\pdfoutput=1

\documentclass[aps,prl,superscriptaddress,showpacs]{revtex4}

\usepackage{graphicx,float}
\usepackage{epsfig}
\usepackage{epstopdf}

\usepackage{amsmath}
\usepackage{amssymb}
\usepackage{bm}
\usepackage{bbm}
\usepackage{dsfont}
\usepackage{bbold}
\usepackage{multirow}
\usepackage{color}
\usepackage[usenames,dvipsnames]{xcolor}
\usepackage[colorlinks=true,citecolor=blue,urlcolor=blue]{hyperref}

\usepackage{hyperref}
\hypersetup{colorlinks=true,linkcolor=blue,citecolor=blue,urlcolor=blue}

\newcommand{\be}{\begin{equation}}
\newcommand{\ee}{\end{equation}}

\newcommand{\sket}[1]{{\ensuremath{\lvert#1\rangle}}}
\newcommand{\lket}[1]{{\ensuremath{\left\lvert#1\right\rangle}}}
\newcommand{\ket}[1]{\if@display\lket{#1}\else\sket{#1}\fi}

\newcommand{\sbra}[1]{{\ensuremath{\langle#1\rvert}}}
\newcommand{\lbra}[1]{{\ensuremath{\left\langle#1\right\rvert}}}
\newcommand{\bra}[1]{\if@display\lbra{#1}\else\sbra{#1}\fi}

\newcommand{\sbraket}[2]{{\ensuremath{\langle#1\rvert#2\rangle}}}
\newcommand{\lbraket}[2]{{\ensuremath{\left\langle#1\!\left\rvert\vphantom{#1}#2\right.\!\right\rangle}}}
\newcommand{\braket}[2]{\if@display\lbraket{#1}{#2}\else\sbraket{#1}{#2}\fi}

\newcommand{\sketbra}[2]{{\ensuremath{\lvert #1\rangle\!\langle #2\rvert}}}
\newcommand{\lketbra}[2]{{\ensuremath{\left\lvert #1\right\rangle\!\!\left\langle #2\right\rvert}}}
\newcommand{\ketbra}[2]{\if@display\lketbra{#1}{#2}\else\sketbra{#1}{#2}\fi}

\begin{document}


\title{Polarization-independent single-photon switch
based on a fiber-optical Sagnac interferometer for quantum communication networks}


\date{\today}

\author{A. Alarc\'{o}n}
\affiliation{Institutionen f\"or Systemteknik, Link\"opings Universitet, 581 83 Link\"oping, Sweden}
\affiliation{Departamento de Ingenier\'ia El\'ectrica, Universidad de Concepci\'on,160-C Concepci\'on, Chile}

\author{P. Gonz\'alez}
\affiliation{Departamento de F\'{\i}sica, Universidad de Concepci\'on, 160-C Concepci\'on, Chile}
\affiliation{Millennium Institute for Research in Optics, Universidad de Concepci\'on, 160-C Concepci\'on, Chile}

\author{J. Cari\~ne}
\affiliation{Millennium Institute for Research in Optics, Universidad de Concepci\'on, 160-C Concepci\'on, Chile}
\affiliation{Departamento de Ingenier\'ia El\'ectrica, Universidad Cat\'olica de la Sant\'isima Concepci\'on, Chile}

\author{G. Lima}
\affiliation{Departamento de F\'{\i}sica, Universidad de Concepci\'on, 160-C Concepci\'on, Chile}
\affiliation{Millennium Institute for Research in Optics, Universidad de Concepci\'on, 160-C Concepci\'on, Chile}

\author{G. B.~Xavier}
\email{guilherme.b.xavier@liu.se}
\affiliation{Institutionen f\"or Systemteknik, Link\"opings Universitet, 581 83 Link\"oping, Sweden}


\begin{abstract}
An essential component of future quantum networks is an optical switch capable of dynamically routing single-photons. Here we implement such a switch, based on a fiber-optical Sagnac interferometer design. The routing is implemented with a pair of fast electro-optical telecom phase modulators placed inside the Sagnac loop, such that each modulator acts on an orthogonal polarization component of the single-photons, in order to yield polarization-independent capability that is crucial for several applications. We obtain an average extinction ratio of more than 19 dB between both outputs of the switch. Our experiment is built exclusively with commercial off-the-shelf components, thus allowing direct compatibility with current optical communication systems.

\end{abstract}




\maketitle


\section{Introduction}

Quantum information employs individual and entangled quantum systems in order to carry a number of information processing tasks offering an advantage over their classical counterparts \cite{Nielsen_2000}. One major sub-division is called quantum communication, which aims to faithfully transmit photonic quantum states across communication links (optical fibers or free-space channels) between remote parties (typically called Alice and Bob) \cite{Gisin_2007}. One important quantum communication protocol is called quantum key distribution (QKD), whose goal is to remotely generate a shared secret key between Alice and Bob \cite{Lo_2014, Xu_2019, Pirandola_2019}. Its effectiveness has been demonstrated over long distances \cite{Boaron_2018}, which is desirable for practical applications. In the past, most quantum communication experiments have focused on point-to-point applications, and only recently interest has increased over network and multi-user applications, with significant effort focusing on the underlaying communication infrastructure supporting future networks of quantum computers, the so-called Quantum Internet \cite{Wehner_2018}.

As in a standard communication network, routing will be an essential function in order to implement dynamic functionality the single-photons. A direct way to implement single-photon routers with potentially fast response time is through the use of interferometers \cite{Ma_2011,Rambo_2013,svarc,Hall_2011}.
In \cite{Ma_2011} a Mach-Zehnder interferometer (MZI) with a phase modulator in one of its arms was used to route the single-photon on demand to one of its outputs. Single-photon switches with two inputs and two outputs have also been presented based on a MZI design \cite{Rambo_2013}. In \cite{svarc} a coupler based on a MZI was presented as well, where photons can be routed at any splitting ratio as a tunable switch. 
In these papers three routing configurations are presented and in all of them an extra active phase stabilization system is required due to the use of MZI.

Aiming for a more stable design, another configuration employed a Sagnac fiber-optical interferometer \cite{Hall_2011}. In this case the switching was performed by modulation of the Kerr non-linearity provided by an auxiliary external optical pulse. More complex single-photon routers based on one-atom switches \cite{Shomroni_2014} and solid-state quantum memories \cite{Sun_2018} have also been presented.


Single-photon switches have applications in other scenarios as well. For instance, one important QKD protocol is based on the use of entangled photon pairs in order to establish a shared secret key between remote parties (typically called Alice and Bob) \cite{Ekert_1991}. One crucial point is to avoid the presence of practical imperfections or limitations, usually referred to as loopholes \cite{Larsson_2014}, that can compromise the security. A popular implementation is based on energy-time entanglement, where the unpredictability on the emission time of photon pairs leads to an entangled state \cite{Franson_1989}. Energy-time entanglement is well suited for long-distance propagation over optical fibers due to its robustness for decoherence \cite{Marcikic_2004}. One typical loophole present in energy-time implementations is due to the required temporal post-selection procedure \cite{Aerts_1999}. This loophole has been recently resolved both in the continuous \cite{Lima_2010, Cuevas_2013, Carvacho_2015}, and pulsed pump respectively \cite{Vedovato_2018}. In the specific case of the pulsed pump (referred to as time-bin entanglement), high-speed optical switches are needed, and in \cite{Vedovato_2018} interferometric switches based on a MZI configuration were employed. Optical switches are also needed in some QKD schemes where fast decoding for path-encoded qubits is needed \cite{Gonzalez_2015}.

In this paper we propose and experimentally demonstrate a high-speed optical switch based on a fiber-optical Sagnac interferometer. Like in \cite{Hall_2011}, our proposal allows intrinsic phase stability due to the Sagnac design and thus does not require additional control systems, an issue that is present in implementations based on an MZI. As an improvement over previous designs, our new configuration uses fast telecom electro-optic phase modulators placed inside the interferometer to provide the required relative phase difference. Since we only rely on off-the-shelf components, we expect our work to be directly applicable to quantum networks and optical network operating within the telecom infrastructure. As a second important improvement over previous designs, we provide a Sagnac interferometer with polarization-independent switching capabilities even though telecom modulators are only able to modulate the phase for one defined polarization. This is possible by resorting to an innovative design for placing the modulators inside the Sagnac loop. The fact that the phases can be properly applied independently of the polarization of the incident photon, opens new possibilities for adopting our switch in quantum information protocols relying on polarization entanglement, which is arguably the main resource of quantum non-locality. For instance, our Sagnac loop can be used for controlling the degree of entanglement of polarization entangled states \cite{Glima09}, for modern self-testing protocols of quantum measurements \cite{Glima16}, and for quantum random number generation based on quantum non-locality \cite{Glima18}.

\section{Experimental description}

In our work we occupy the stability properties of a Sagnac interferometer. A Sagnac interferometer consists of a beamsplitter whose two outputs are connected together, such that the light beams propagate through the same path over the two opposing directions, yielding passive and intrinsic stability to the interferometer \cite{Culshaw_2006}. The probability that a single-photon exits through either output port depends on the relative phase-shift imposed between the two counter-propagating paths. If the interferometer is subjected to slow perturbations (compared to the total propagation time between input and output), then the probability to come out from the input port is always unity, since both paths undergo the same phase shift \cite{Culshaw_2006}. For this reason, the Sagnac is stable against slow phase perturbations. 

\begin{figure}[h!]
\centering\includegraphics[width=12cm]{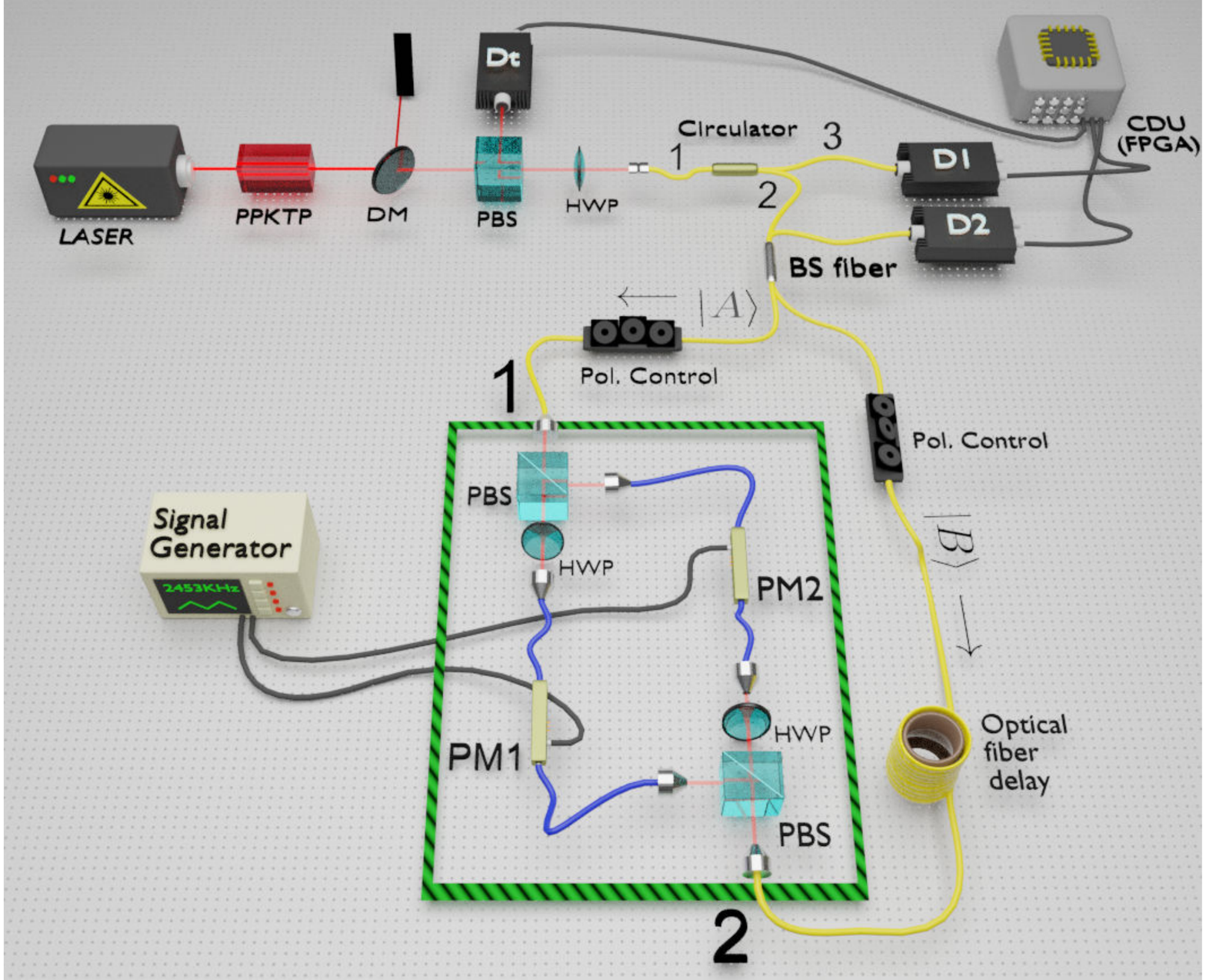}
\caption{Experimental setup for the polarization-independent single-photon switch. In the area inside the green border is the Mach-Zehnder structure housing both phase modulators to yield polarization independency. Please see text for details. \label{Fig1} }
\end{figure}

The experimental setup is depicted in Fig. \ref{Fig1}. We take advantage of a Sagnac configuration while employing two phase modulators inside the Sagnac loop in order to yield polarization independence with respect to the input state. We generate single-photon pairs through the process of spontaneous parametric down-conversion (SPDC) in a non-linear periodically poled potassium titanyl phosphate (PPKTP) waveguide crystal \cite{Fiorentino_2007}. The crystal (ADVR inc.) is phase-matched to create degenerate orthogonally-polarized (type-II) 1546 nm photon pairs when pumped with laser light with a wavelength of 773 nm. The waveguide is designed to only support a single propagation mode for the pump light and the down-converted photons. Light from an external cavity tunable laser is coupled to the waveguide after focusing with an 11 mm focal length aspheric lens, followed by an identical one at the output collimating the wavefront of the down converted photons. A dichroic mirror is used to remove excess pump light before the photons are deterministically split at a polarizing-beam splitter (PBS). At each output port of the PBS, the single-photons are coupled to single-mode fibers through multi-axis translation stages. 

The reflected output from the PBS following the crystal is connected to a free-running mode InGaAs single-photon detector module (D$_\mathrm{t}$), with 15\% overall detection efficiency (IdQuantique id220). The output from this detector is used to trigger the two other single-photon detectors D$_1$ and D$_2$ (IdQuantique id210, working in gated mode, also 15\% detection efficiency) placed at the outputs of the Sagnac interferometer. The Sagnac is built from a 50:50 fiber beamsplitter (BS), whose outputs create the two paths $|A \rangle$ and $|B \rangle$ indicated in Fig.\ref{Fig1}. Inside the Sagnac a Mach-Zehnder-like structure (MZs) with polarizing beam splitters (PBS) placed at its input and output is constructed to house the orthogonally placed phase modulators (PM). The paths $|A \rangle$ and $|B \rangle$ reach the MZs from opposite directions, as seen in Fig. \ref{Fig1}.

Both PBSs are free-space units mounted in a compact module with input and output integrated coupling lenses. A half-wave plate (HWP) placed in the same PBS module (shown in the inset in Fig. \ref{Fig1}) is used to ensure the correct polarization state (vertical) enters one of the fiber-pigtailed lithium niobate electro-optic phase modulators (PM) with 10 GHz bandwidth. The input/output fibers from both PMs are polarization maintaining. The MZs is therefore employed to decompose each $|A \rangle$ and $|B \rangle$ paths into orthogonal polarization components, with one of these rotated by 90$^{\circ}$ by the HWP. This arrangement is employed as standard commercial PMs only act upon one polarization component, in order to ensure the polarization insensitivity of our setup. 

\begin{figure}[h!]
\centering\includegraphics[width=12cm]{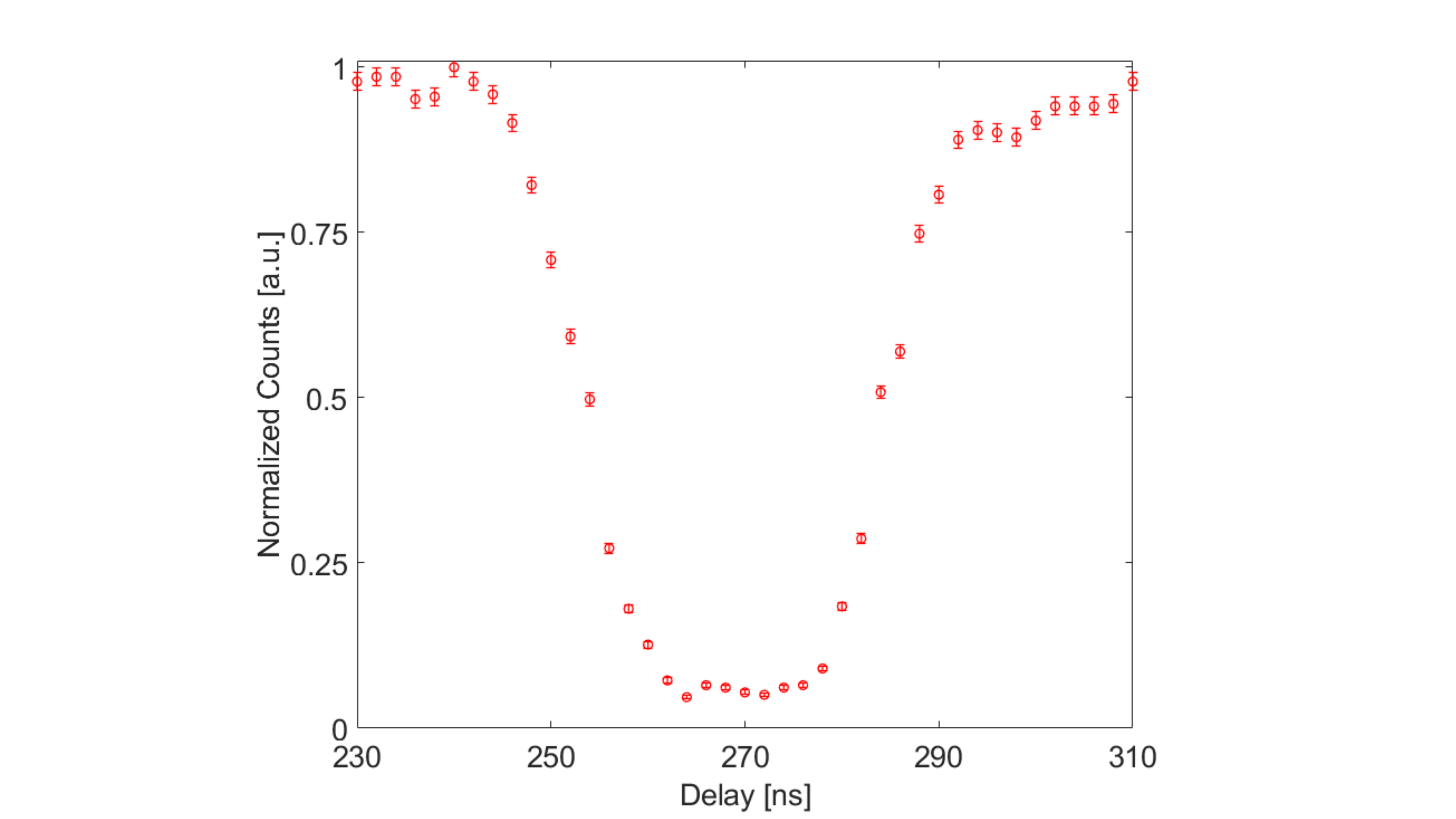}
\caption{Coincident single-photon detections between D$_t$ and D$_1$ as a function of the relative delay imposed on the driving pulses applied to the phase modulators. The integration time is 10 s, and the error bars show the standard deviation. \label{Fig2}}
\end{figure}

Following the MZs outputs, both $|A \rangle$ and $|B \rangle$ components propagate back to the fiber BS. There is a 100 m optical fiber delay between input  $|B \rangle$  and the MZs, which is used to ensure that the modulation imposed at the PMs only acts upon one of the propagating directions in the Sagnac. After recombination at the BS, the single-photon can propagate to either D$_1$, following the optical circulator, or D$_2 $. Both modulators are driven by a single signal generator producing 32 ns wide pulses. The generator is triggered by D$_\textrm{t}$, with electronic delays used to appropriately synchronize the timing. The number of detections per second is processed by a field programmable gate array (FPGA) coincidence detection unit (CDU). Fig. \ref{Fig2} shows the rate of coincident detections at D$_1$, as a function of the delay imposed on the electrical pulse driving the PMs.  

We assume an arbitrary input polarization state at the Sagnac's input of the form  $|\psi\rangle =\alpha|H\rangle + \beta|V\rangle$ , where $|H\rangle (|V\rangle)$, is the quantum state representation for the horizontal (vertical) polarization state respectively and $\alpha$ and $\beta$ are complex coefficients such that $|\alpha|^2 + |\beta|^2 = 1$. After the input BS we have that the state of the photon is $|\psi\rangle=1/\sqrt{2}\left((|A\rangle + i|B\rangle)\otimes(\alpha|H\rangle + \beta|V\rangle)\right)$, where $i$ is the relative phase shift between the transmitted and reflected ports of the BS. Manual polarization controllers inside the Sagnac loop are used to ensure that the both orthogonal polarization components from the input state always propagate through the same path in the MZs, i.e. the $|H\rangle$ component propagating through $|A\rangle$ takes the path containing PM$_1$ in the MZs, while the same $|H\rangle$ component injected in the opposite $|B\rangle$ path is routed to the same PM (since the manual polarization controller switched it to $|V\rangle$ before the PBS input). This is necessary to benefit from the inherent phase stability of the Sagnac. At the end of the interferometer, we can consider $|D_1\rangle$  and $|D_2\rangle$ as the two possible paths (outputs) that $|A\rangle$ and $|B\rangle$ could take after being combined in the beam splitter, the final state of the system is:

\begin{align}\nonumber
|\psi\rangle=\frac{1}{2}\left[\{i(e^{i\phi}+1)|D_{1}\rangle+(e^{i\phi}-1)|D_{2}\rangle\}\otimes\{\alpha|H\rangle+e^{iKl}\beta|V\rangle\}\right] 
\end{align}
where $\phi$ is the phase shift provided by each PM (Since we employed a single generator to drive both PMs, the phase shift applied to each modulator, is the same and equal to $\phi$), $K$ is the propagation constant and $l$ the length difference between the two arms of the MZs, which is guaranteed to be longer than the short-coherence length of the single-photons ($\sim $ 1 mm). We can then write the probability that a single-photon is detected at either output of the Sagnac loop as:

\begin{equation}
|{{\big\langle \textrm{D}_{1}|\psi \rangle }|}^2 = {cos}^2\left(\frac{\phi}{2}\right) \label{eq3}
\end{equation}
\begin{equation}
|{{ \langle \textrm{D}_{2}|\psi \rangle }|} ^2 = {sin}^2\left(\frac{\phi}{2}\right) \label{eq4}
\end{equation}

Therefore the outputs are complementary as a function of $\phi$, as expected for an interferometer. Such a scheme can then be used as an optical switch for photonic quantum states, since the single-photon output probability depends on the applied phase shift $\phi$.

\section{Results}

\begin{figure}[ht]
\centering\includegraphics[width=15cm]{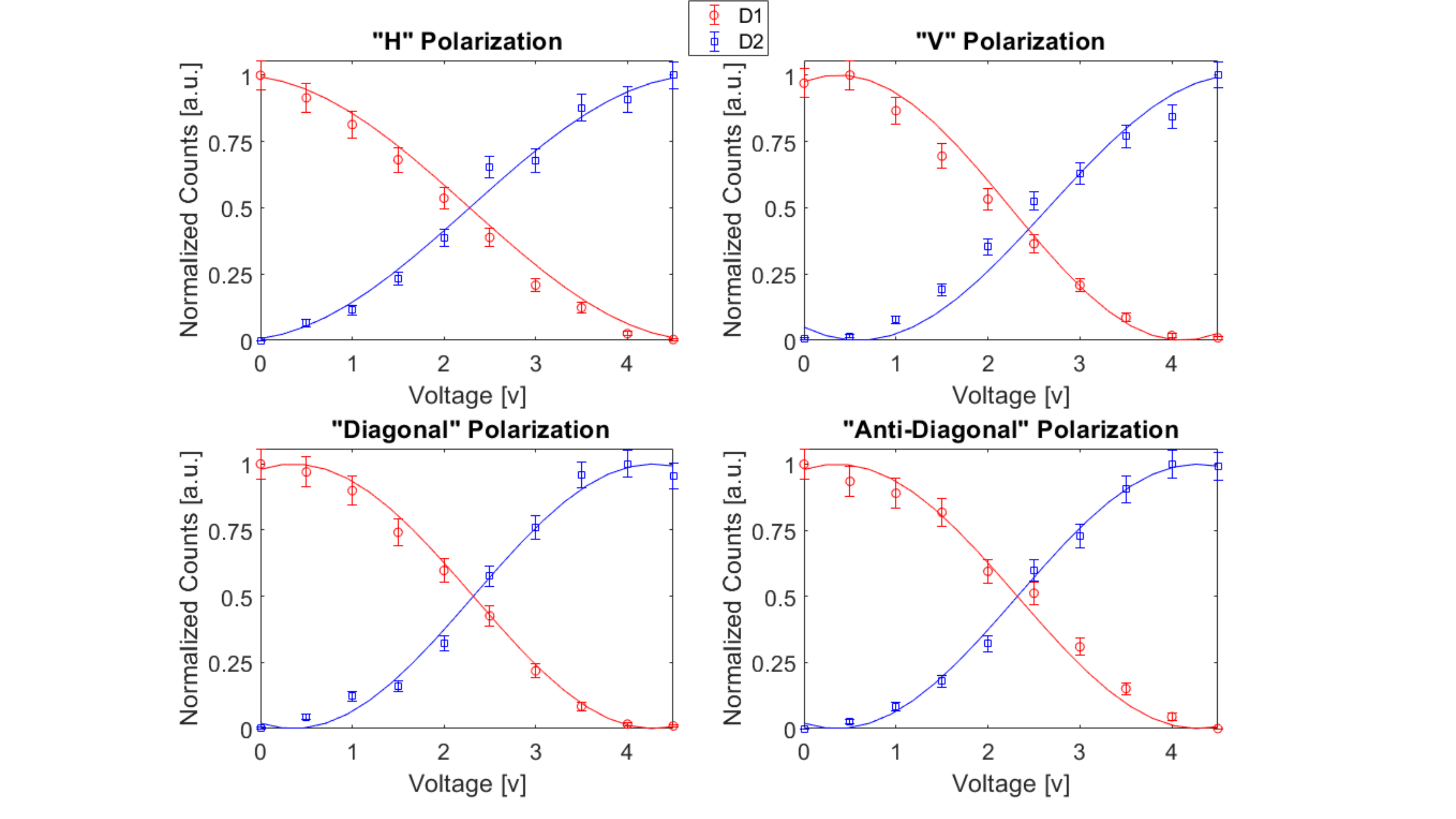}
\caption{Experimental results for four input polarization states: $|H\rangle$ (horizontal),$|V\rangle$ (vertical), $|D\rangle$ (diagonal) and $|A\rangle$ (antidiagonal), as a function of applied voltage to the phase modulators. The data points represent coincident detections at D$_1$ and D$_2$ , both triggered by D$_\textrm{t}$. Error bars represent one standard deviation arising from 20 repeated measurements (10 s integration time per measurement) for each voltage. \label{Fig3}}
\end{figure}

We experimentally demonstrate the scheme measuring the coincidence rate between triggering detector D$_\textrm{t}$ and the two output detectors D$_{1}$ and D$_{2}$, as a function of the relative phase $\phi$, which is proportional to the driving voltage applied to the PMs. This is done initially for an input horizontal polarization-state, set through the HWP placed before the circulator's input (Fig. \ref{Fig1}). For each applied voltage we take 20 measurements of 10 s each, plotting the results in Fig. \ref{Fig3}, with the error bars representing one standard deviation. The x-axis in Fig. \ref{Fig3} corresponds to the voltage applied to the PMs and it is proportional to modifying the relative phase $\phi$ of the Sagnac. From equations (\ref{eq3}) and (\ref{eq4}), we can see that for $\phi = 0^\circ$, the probability of detecting a photon is maximum in D$_1$ and minimum in D$_2$, with the opposite when $\phi = 180^\circ$. From Fig. \ref{Fig3} approximately 4 V are needed for a $\pi$ phase shift. We  demonstrate the polarization-independent character of the setup by repeating the measurement for three other input states: vertical, diagonal  $|\textrm{D}\rangle =\tfrac{1}{\sqrt{2}}\left(|H\rangle + |V\rangle \right)$ and the anti-diagonal state $|\textrm{A}\rangle =\tfrac{1}{\sqrt{2}}\left(|H\rangle - |V\rangle \right)$, obtaining very similar results (Fig. \ref{Fig3}). The total loss in the experiment from the input of the Sagnac, to the single-photon detectors is approximately 5 dB, limited mainly by the insertion loss of the phase modulators.
 
A typical figure of merit of an interferometer is given by the visibility V = $(\textrm{C}_{1} - \textrm{C}_{2}) / (\textrm{C}_{1} + \textrm{C}_{2})$, where C$_1$ and C$_2$ are the number of counts at detectors D$_1$ and D$_2$, when triggered by D$_\textrm{t}$, respectively for the same integration time. We obtain an average visibility over all 4 input polarization states of 97.63 $\pm$ 0.21\% without subtracting accidental counts, leading to an average extinction ratio of 19.21 dB. We obtain 25.1 detections/s on average for all input states in one output, when the applied voltage is adjusted such that constructive interference is obtained at that output.

One consequence of the internal MZs is that it is subjected to relative phase fluctuations between each arm due to thermal variations. While it does not affect the relative phase set between the arms "A" and "B", and thus the Sagnac switching capability, it has the effect that a polarization state that propagates through both arms (in other words any state except $|H\rangle$ or $|V\rangle$) will undergo a polarization rotation at the end of the Sagnac loop. Nonetheless, since thermal fluctuations represent slowing varying polarization rotations, the output polarization states at either port of the switch can easily be compensated with varying HWPs \cite{Xavier_2008}. This effect can be further mitigated by thermally insulating the MZs.

\section{Conclusions}

We have experimentally demonstrated a polarization independent single-photon optical switch that is capable of deterministically routing two different spatial output single-photon modes, while taking advantage of the intrinsic stability of the Sagnac interferometer. The switching speed was only limited by the electronics and detection hardware, but it is fully capable to operate on the GHz range. Our work presents a new design of fiber-optical Sagnac interferometer that has many potential applications in quantum information science. Furthermore, our design can also be used in classical optical networks with the benefit that it works independently of the arrival state of polarization of a light pulse \cite{Agrawal}. Finally our work presents an alternative to previously used switches, since it also has the extra advantage of being fully compatible with optical network hardware, as it uses off-the-shelf components in the 1550 nm telecom window, which support ultra-fast switching speeds.

\section*{Acknowledgments}
We acknowledge Felipe Toledo for experimental assistance and Daniel Mart\'inez, Esteban S. G\'omez and Niklas Johansson for valuable discussions. G. X. acknowledges Ceniit Link\"{o}ping University, the Swedish Research Council (VR 2017-04470) and QuantERA grant SECRET (VR grant no. 2019-00392) for financial support. A. A. acknowledges financial support from the Knut and Alice Wallenberg Foundation through the Wallenberg Center for Quantum Technology (WACQT). G.L. was supported by Fondo Nacional de Desarrollo Cient\'{i}fico y Tecnol\'{o}gico (FONDECYT) (1200859) and Millennium Institute for Research in Optics. P.G-G. acknowledges support from ANID FONDECYT/POSTDOCTORADO/N$^{\circ}$ Proyecto 3200820. J.C. acknowledges  support  from ANID/REC/PAI77190088.  
%


\end{document}